\documentclass[11pt,a4paper,pdftex]{article}
\usepackage{amsmath, amscd, amssymb, amsthm, latexsym, stmaryrd
}
\usepackage{float}

 \usepackage{graphicx}
\usepackage[colorlinks=true,linkcolor=red]{hyperref}
\usepackage{hyperref}

\newcommand{\N}{\mathbb{N}}

\newcommand{\Z}{\mathbb{Z}}
\newcommand{\nn}{N}
\newcommand{\kk}{K}
\newcommand{\Ss}{s}%
\newtheorem{theorem}{Theorem}

\newtheorem{lemma}[theorem]{Lemma}

\usepackage{tikz,pgfplots}

\usetikzlibrary{arrows,automata,plotmarks,decorations,positioning}
\tikzstyle{post}=[->,shorten >=1pt,>=stealth']
\tikzstyle{every initial by arrow}=[initial text={},initial distance=1em,post]
\tikzstyle{transition}= [post,shorten >=1pt,node distance=2cm, inner sep=2pt,bend angle=20]

\begin{document}

\title{Message complexity on unary multiautomata systems}

\author{ Christian Choffrut\\
IRIF (UMR 8243), \\
CNRS and Universit\'e Paris Cit\'{e},  
France\\
\tt{Christian.Choffrut@irif.fr}
}
\date{}
\maketitle

\date{}
\maketitle

\begin{abstract}
Finitely many two-way automata work independently and synchronously on a unary input. Some of their states are broadcasting, i.e., 
dispatched to all other automata.  At each step of the computation, each automaton
changes state and moves right, left or stay in place according to the current state and the possible messages dispatched. 
The input is recognized if the following occurs: starting from the initial configuration (the heads of all automata are positioned to the left end of the tape) one automaton reaches a final state when its head is positioned to the right end 
of the tape. 
We show that if the number of messages sent during the computation is bounded by some integer which is independent of 
the length of the input, then the language recognized is regular, 
\end{abstract}

{\bf Key words}: multiautomata, two-way automata, unary language, communication complexity, 

\section{Introduction}

We investigate a model introduced by 
Tomasz Jurdzinski in his thesis, see also \cite{JuKu2001}.  A fixed number 
of finite (deterministic or not, one-way or two-way) automta work independently and synchronously on the same
 input  guarded by two endmarkers, the left and right markers $\triangleright$  and $\triangleleft$, 
to prevent
their heads  to fall off. They have the possibility  of broadcasting or not a message at each time unit. According to their current state and the 
possible messages received from the other automata, they  change state and direction.  The resource in this setting is the number of communication steps (we will say simply messages by considering several messages sent simultaneously as a unique message), occurring during the 
computation expressed as a function $f(\nn)$  of the length $\nn$  of the input, the \emph{message complexity}. There are many variants: deterministic/nondeterministic, one-way/two-way, size of the alphabet 
 (essentially and as usual unary/nonunary), size of the memory. Caution: the model is not to be confused 
  with $k$ head  one- or -two-way finite automata. Indeed, in this latter case the memory is central which means that  the device can be viewed as  
a multiautomaton  exchanging constantly  messages.  

Formally, Tomasz Jurdzinski  considers a system $\+M=(\+A_{1}, \ldots, \+A_{n})$ of finite automata, also called \emph{multiautomaton},
  with state set $Q$. The individual automaton $\+A_{i}$ in state $s$ and reading a letter $a$ may broadcast or not a message $\mu_{i}(\Ss,a)$ (when no message is sent, this message  is interpreted as nil). In state $\Ss$ upon
reception of  the vector of messages $(m_{1}, \ldots, m_{n})$   the next state of automaton $\+A_{i}$ is defined by the condition
\begin{equation}
\label{eq:transition-function}
(\Ss',d)\in  \delta_{i} (\Ss, a, (m_{1}, \ldots, m_{n}))\quad m_{j}= \mu_{i}(\Ss_{j},a) \text{ or } \texttt{nil} \quad j=1, \ldots, n
\end{equation}
where $\delta_{i}$ is the \emph{transition relation},  $\Ss'$ is a next state and the head moves from the current position $p$ to 
the next position $p+d$ with $d=-1,0,+1$. A configuration is an $n$-tuple
\begin{equation}
\label{eq:configurations}
((s_{1},p_{1},a_{1}), \ldots, (s_{n},p_{n},a_{n}))
\end{equation}
where $s_{i}$ is the current state of automaton $\+A_{i}$, $p_{i}$
the position of its head on the input and $a_{i}$ the letter scanned. 
The automata start in an initial configuration, i.e., in some initial state of their own  (which is the reason why 
this is no loss of generality to assume the same set of states $Q$) and with their head  positioned on the left endmarker. 
A run is a sequence of 
configurations obeying the instructions of the transition relation. 
An input $w$ is accepted if a starting from an initial configuration, a run reaches a configuration where some  predetermined automaton (say 
$\+A_{1}$ without loss of generality) enters a specific \emph{final state}, i.e., 
the first component of the configuration \ref{eq:configurations} is of the form $(s_{\text{fin}}, |w|+1, \triangleleft)$.
The system is allowed to send an amount of messages up to a certain value depending on
the length of the input.  For the one-way version the moves are restricted to $d=0,1$.
Tomasz Jurdzinski and
                  Miroslaw Kutylowski  improved on  results in Jurzinski's thesis: there exists no language recognizable
by a one-way deterministic system with message complexity  between $\omega(1)$ and $o(\log \nn)$,
\cite[Thm 1]{JuKu2001}  
and  for some  constant $c$ there exists no language recognizable
by a two-way deterministic multiautomaton with message complexity  between $\omega(1)$ and $o((\log\log\log \nn)^c)$,
\cite[Thm 2]{JuKu2001}.  
%

\section{Premiminaries}

Since  we   deal with unary inputs we make the following convention: we ignore the
 second  component of the transition function in the expression \ref{eq:transition-function}  and the third component 
in expression \ref{eq:configurations} when it is understood that this is the unary input $a$
and not the endmarkers $\triangleright, \triangleleft$.
However 
the endmarkers are explicit when  scanning one of them. 
Also, when considering a single automaton we drop the third component of the transition function
and the index $i$. 
  The nonempty messages of
\ref{eq:transition-function} are identified with the current states of the automata.

 Consider a two-way automaton moving on a two-way infinite unary tape. 
For all states $t,t'$,  we  define $\epsilon(t)=d$ if $(t',d)=  \delta(t)$.
The \emph{basic sequence} of state $s$ is the sequence of successive states  of transitions determined by $\delta$
\begin{equation}
\label{eq:basic-sequence}
s=s_{0}, s_{1}, \ldots, s_{\ell}, \ldots, s_{k}=s_{\ell}  
\end{equation}
where all elements are different except $ s_{k}$ and $s_{\ell}$.
We set  $\lambda_{s}(s_{0})=0$ and with every $i>0$, we
set $\lambda_{s}(s_{i})= \epsilon(s_{0})+ \cdots + \epsilon(s_{i-1})$. 
 In state $s$ the automaton  is 
\emph{virtually moving to the right} (resp. \emph{to the left}) if  
$\lambda_{s}(s_{k-1}) -\lambda_{s}(s_{\ell})>0$ (resp. $\lambda_{s}(s_{k-1}) -\lambda_{s}(s_{\ell})<0$). 
It is \emph{virtually motionless} in all other cases. The \emph{amplitude} of $s$ is the integer 
$\max \{\lambda_{s}(s_{i})| i\leq  k\} -\min \{\lambda_{s}(s_{i})| i\leq k\} $. When the automaton 
scans a finite tape, it may happen, because of the ``initial mess'',  that it falls off  the two  ends of 
the tape before entering the loop which would make it move to the right or to the left.

\bigskip The multiautomaton is composed of $n$ individual complete unary (except for the two endmarkers) 
deterministic automata $\+A_{i}$ having
 disjoint  state sets  $Q_{i}$, $i=1, \ldots, n$,  each 
 with a unique  initial state and some final states. 
The set of {\em broadcasting} and {\em final states} are denoted $B_{i}$ and $F_{i}$.

The automata work on the same input $\{0, 1, \ldots, \nn, \nn+1\}$, $\nn\in \N$ where $0$ and $\nn+1$ are the 
positions of the two endmarkers.
 A \emph{global state} is an $n$-tuple of states 
$\sigma\in Q_{1}\times \cdots \times Q_{n}$. A \emph{global position}
  is a vector  $\pi=(\pi_{1}, \ldots, \pi_{n})\in \{0, \ldots,  \nn +1\}^n$ of the positions of the heads over the tape.
A \emph{configuration} is a pair $(\sigma, \pi)$ where $\pi$ is a global position
and $\sigma$ is a global state.
A global  position $\pi$ is  \emph{hitting} if  $\pi_{i}=0$ or $\nn+1$ for some $1\leq  i \leq n$. It is
 \emph{broadcasting} if for some $1\leq i\leq n$ the state of automaton  $\+A_{i}$ is in
$B_{i}$. We speak similarly of hitting and broadcasting configurations in the obvious  way. The
pair $(\sigma, \pi)$ is the  \emph{initial configuration} 
if for all $i=1, \ldots, n$ $\pi_{i}=0$  and $\sigma_{i}$ is the initial state of $\+A_{i}$.
It is a \emph{final configuration} if  $\pi_{1}=\nn+1$  and if $\sigma_{1}$ is a final state of $\+A_{1}$. 
Furthermore, since exactly $M$ messages are allowed, we may assume without loss of generality that each automaton 
is provided with a counter that prevents it from sending more than $M$ messages.

 {\em In order to avoid the hectic behavior of the automata on small inputs,  I
consider without loss of generality only  inputs  greater than  the amplitudes of all states of all automata.
In particular the two endmarkers are not  both reached  in the basic sequence \ref{eq:basic-sequence}.
The expression ``sufficiently large'' in the sequel refers to this convention, either for individual automata or for
the collection of  automata $\+A_{i}, i=1, \ldots n$ depending on the context.}

The dynamic of the multiautomaton is the sequence of configurations as time goes on. It is 
defined as follows. Given 
the current global state $\sigma$ and the current global position $\pi$, 
the next global  position is the vector  $\pi'$
where for $i=1, \ldots, n$ we have $\pi'_{i}= \pi_{i} +  d$ 
with  $(\sigma'_{i},d)=  \delta_{i} (\sigma_{i}, x, (m_{1}, \ldots, m_{n}))$, $x\in \{a, \triangleleft, \triangleright\}$
and $m_{j}=\texttt{nil}$ or $\sigma_{j}$, $j=1, \ldots, n$. The input is recognized if starting from the initial configuration, the multiautomaton reaches eventually some final configuration. %

\begin{theorem} 
\label{th:finite-number-of-messages} For every integer $M$, the unary languages recognized by deterministic two-way multiautomata exchanging
$M$ messages,   are regular.
\end{theorem}

 The proof consists of showing that there exists a Presburger  formula with the unique free variable $\nn$  asserting that the input of length $\nn$
 is recognized by the multiautomaton. We recall that  the Presburger formulas
  are the first-order formulas of the structure $\langle \Z: 0, (x,y) \rightarrow x+y, x\leq y \rangle$.
  In other words, they
  are  obtained from the constant $0$, the basic predicate $z=x+y$ and $x\leq y$
by using the logical connectives (disjunction,  conjunction and  negation) and the universal and existential quantifiers,
\cite{presburger}. 
This  is indeed sufficient because a subset of the integers $\N$ is regular if and only if it satisfies a Presburger formula, see e.g., \cite[Thm 1.3]{GS}.
We use (hopefully natural) shorthands in order to keep the formulas readable, such as multiplication by integer constants, introduction of constants
that encode for the states, for subsets for indices in $\{1, \ldots, n\}$ etc\ldots
We assume $M>0$. Indeed, if $M=0$, only one automaton matters ($\+A_{1}$ by convention) and it is known that for arbitrary alphabets the languages recognized by two-way finite automata are regular, cf. \cite{RabSco}. 

The question might seem trivial. Indeed, between two consecutive broadcasting steps,
 the automata work
independently. When one of them reaches a broadcasting state the computation performed is exactly that
which  it would have 
performed if it had been left alone, in which case it ``would have recognized 
a regular language''.  Since there are finitely many messages it looks as if would suffice to ``compose''
these computations in which case we would end up with a Boolean combination of regular languages. 
However, the positions of the heads at the end of a communication step (and the beginning of a new communication step) 
are arbitrary and we cannot record all these possible positions. This difficulty is overcome as follows. Assume the computation 
(between two consecutive messages) begins in some global state $\sigma$ and ends in some global state $\tau$;
assume
further that the distribution of the possible global positions at the beginning of the step satisfy some 
first-order formula $f(\nn, \pi_{1}, \ldots, \pi_{n})$ (for example with  two automata on input $\nn$, the formula 
$\pi_{1}= \lceil \frac{\nn}{2}\rceil +1 \wedge \pi_{2}=3$). Then the crux consists of showing 
that the distribution of the possible global positions satisfy 
another first-order formula depending uniquely on $\sigma, \tau$ and $f$, see lemma \ref{le:new-recognition}
for a more detailed presentation. The conclusion is met by remarking that
 there are finitely many global states, finitely many  broadcasting steps and 
 the distribution of the possible initial positions is the clearly first-order formula: $ \pi_{1}= \cdots = \pi_{n}=0$.

\section{Dynamic of a single automaton}

First, consider the dynamic of a single automaton in absence of broadcast. The  next  result is folklore. It says that when starting from 
an endpoint, either the head returns to the same endpoint without visiting the opposite endpoint or it reaches the opposite endpoint 
without returning to the initial endpoint or it gets stuck in the middle.

\begin{lemma}
 \label{le:take-off}  Let  $\nn$ be sufficiently large. Given a configuration $(s,0)$ (resp. $\nn+1$) with the parameters $\ell$
 and $k$ 
 of expression \ref{eq:basic-sequence}, one of the following assertions is true.
 
 \begin{itemize}
 
 \item there exists a time $T\leq k+\frac{k^2}{4}$   such that the head returns to the left (resp. right) end of the tape 
 after $T$ units of time
 
 \item the exists three integers $p$ and $T_{1}, T_{2}\leq k$ 
 such that the head reaches position $p< k$  (resp.$\nn-p<k$) after $T_{1}$ units of time and oscillate around $p$ with period $T_{2}$
 without ever visiting the positions $0$ and $\nn+1$.
 
 \item the head reaches eventually position $\nn+1$ (resp. $0$)  without ever returning to position $0$ (resp. $\nn+1$).
 
 \end{itemize}
 
\end{lemma}

\begin{proof} It suffices to treat the case where the initial configuration is at position $0$.
Consider expression \ref{eq:basic-sequence}
and assume  
$c=\lambda_{s}(s_{k-1}) - \lambda_{s}(s_{\ell})\geq 0$. If $\lambda_{s}(s_{i})=0$ for some 
$i<k$ we are done with $T\leq k$. Otherwise if $c=0$ then the head oscillates around the position $\lambda_{s}(s_{\ell})$
with period $k-\ell$. If $c>0$ the head moves right until is hits the right end of the tape. If $c<0$
then it hits the left end of the tape at time  $T\leq \ell + \lceil \frac{\ell}{|c|}\rceil (k-\ell)\leq \ell + \ell \cdot(k-\ell)\leq k+\frac{k^2}{4}$

\end{proof}

\begin{lemma} 
 \label{le:norebound} Given a subset $S\subseteq Q$ and two  positions $0\leq p,p'\leq \nn+1$,
 there exists an effective  first-order formula $\text{\em Reach}_{S}(N,s,s',$ $p,p',T)$ that expresses the fact that starting in configuration $(s,p)$
the automaton reaches configuration $(s',p')$ at time $T$ without ever visiting
a position $0$ or $\nn+1$ in the mean time and without visiting any state in $S$ except 
possibly at time $T$. 
\end{lemma}


\begin{proof}
We assume first $p\not\in \{0,\nn+1\}$.
Consider the basic sequence \ref{eq:basic-sequence}
and assume  
$c=\lambda_{s}(s_{k-1}) - \lambda_{s}(s_{\ell})\geq 0$.  Let $i$ be the smallest integer,
if it exists,  such that $s_{i}\in S$ or  $p+\lambda_{s}(s_{i})=0$ or 
$p+\lambda_{s}(s_{i})=\nn+1$. Then 
for  $j=0, \ldots, i$  we have
\begin{equation}
\label{eq:c1}
 (T=j) \wedge (p'=p+ \lambda_{s}(s_{j} ))
\end{equation}

Now we may assume 
that no state   in $S$ is visited during the run from $p$ to $p'$. 
Let $\ell \leq m<k$ be the smallest integer $u$
satisfying 
$s_{u}=\arg \min \{\lambda_{s}(s_{v}) \mid \ell \leq v<k\}$.
For each of the finitely many cases $0\leq i\leq m$,  the following holds
$$
p'=p+\lambda_{s}(s_{i}) \wedge (T=i)
$$
Thus, we may assume that the basic sequence of $s$ satisfies $\ell=0$ ($s$ belongs to a cycle) and that 
$\min\{\lambda_{s}(s_{i})\mid 0\leq i<k \}=0$ (the head never moves to the left of the initial position). 
Now, consider configuration $(s',p')$ where $s'=s_{i}$ for some   $0\leq i <k$. There exists a unique integer $H$ such that 
   $p'=Hp+ \lambda_{s}(s_{i})$.
Consider the longest subsequence of states visited while at position $p'$, say $s_{i_{1}}, \cdots, s_{i_{m}}$,
such that
 $\max \{ \lambda_{s}(s_{j})| i_{1}\leq j\leq i_{m}\} <N+1-p'$ (the run stays inside the input), see Figure \ref{fig:case1}. Observe that this maximum  takes 
on  only finitely many values and  that we have for all $i_{r}$,  $r=1,  \ldots, m$

 \begin{equation}
\label{eq:c2} \exists H (T=H\cdot k + i_{r}) \wedge (p'=p+ H\cdot c  + \lambda_{s}(s_{i_{r}}))
\end{equation}

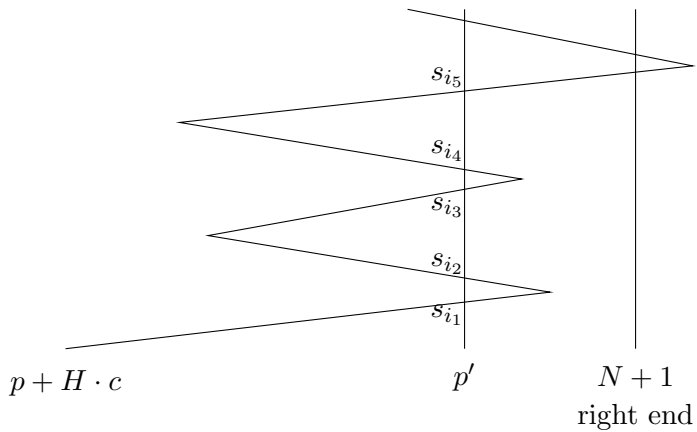
\begin{figure}[H]
\begin{center}
\begin{tikzpicture}[scale=0.75]
\draw (7,0) -- (7,6);
\draw (10,0) -- (10,6);
\draw (0,0) -- (8.5,1) -- (2.5,2) -- (8,3) --(2,4)--(11,5)--(6,6);
\node at (7,-0.5) {$p'$};
\node at (10,-0.5) {$\nn+1$};
\node at (10,-1.2) {right end};
\node at (0,-0.6) {$p+H\cdot c$};
\node at (6.7,0.6) {$s_{i_{1}}$};
\node at (6.7,1.5) {$s_{i_{2}}$};
\node at (6.7,2.5) {$s_{i_{3}}$};
\node at (6.7,3.5) {$s_{i_{4}}$};
\node at (6.7,4.8) {$s_{i_{5}}$};
%
 \end{tikzpicture}
\end{center}
\caption{$\lambda_{s}(s_{i_{1}})= \lambda_{s}(s_{i_{2}})= \lambda_{s}(s_{i_{3}})= \lambda_{s}(s_{i_{4}})= \lambda_{s}(s_{i_{5}})$. In the loop, the head passes $5$ times at position $p'$ (not $6$ times because it falls off before)}
\label{fig:case1}
\end{figure}

It remains to consider several special cases. Consider $p=0$ and $(t,d)=\delta(s,\triangleright)$.
If $d=-1$  no configuration is 
reachable from$(s,0)$. If $d=0$
the unique configuration reachable under  the condition of the lemma 
is $(s',p')= (t,0)$ in time $T=1$. If $d=1$ then the configuration $(s',p')$ is reachable from $(s,p)$ in time $T$
under the condition of the lemma if and only if it is reachable from $(t,1)$ in time $T-1$. 
The case $p=\nn+1 $ can be treated similarly.

The previous discussion assumes the hypothesis $c=\lambda_{s}(s_{k-1}) - \lambda_{s}s_{(\ell})\geq 0$
but can trivially be adapted to the hypothesis $c\leq 0$.
\end{proof}

A \emph{traversal} is a run between two   configurations $(s,0)$ and $(s',\nn+1)$
(resp. $(s,\nn+1)$  and $(s',0)$)
without rebound on the right (resp. left) endmarker. It may rebound on  $(s,0)$ 
(resp. $(s,\nn+1)$  but since $\nn$ is sufficiently large, the number of  rebounds  is finite
and independent on the length of the input. We use the terms right traversal and left traversal with the natural meaning.

\begin{lemma}
 \label{le:traversal} Given two hitting configurations $(s,0)$ and $(s',\nn+1)$
(resp. $(s,\nn+1)$  and $(s',0)$),  there exists an effective  first-order formula
 $\text{\em Right}_{S,s,s'}(N,T)$ (resp.  $\text{\em Left}_{S,s,s'}(N,T)$)
        that holds if and only if $T$
is the time of the traversal 
  from $(s,0)$ to
$(s',\nn+1)$ (resp.  $(s,\nn+1)$  to $(s',0)$).

\end{lemma}

\begin{proof}
It suffices to consider the case from  $(s,0)$ to 
$(s',\nn+1)$.  Since $\nn$ is sufficiently large, if $(s',\nn+1)$ is reachable from $(s,0)$, by lemma \ref{le:take-off}
 there exists a fixed sequence (independent of $\nn$)
$s_{0}=s, s_{1}, \ldots, s_{q}, s'$ such that the automaton passes successively through 
the configurations $(s_{0},0), (s_{1},0), \ldots, (s_{q},0), (s',\nn+1)$ and no other hitting configurations. 
\begin{equation}
\label{eq:right-left}
\begin{array}{l}
\text{Right}_{S,s,s'}(N,T)\equiv \exists T_{1}, \ldots, T_{q+1}\quad 
 (T=T_{1}+ \ldots, +T_{q+1})\\
  \wedge \text{Reach}_{S}(N,s_{q},s',0,\nn+1,T_{q+1})  \wedge \displaystyle \bigwedge_{1\leq i\leq q} \text{Reach}_{S}(N,s_{i-1},s_{i},0,0,T_{i})
\end{array}
\end{equation}
\end{proof}
%


We introduce  four predicates which specify how the  head hits the end positions successively. For example
the first predicate is true if $T$ is the time to perform a zigzag starting in position $0$, ending in 
positions $\nn+1$ and rebouncing at least once  on the hitting positions.
The other three predicates are interpreted in a similar way. Figures \ref{fig:RR},  \ref{fig:RL},  \ref{fig:LR},  \ref{fig:LL} might help the reader.
The integer $K$ is the maximum number of hits allowed.
\begin{center} \fbox{$2r$ bounces} \end{center} 
$$
\begin{array}{ll}
\text{ RR}_{S, s_{0},  s_{1}, \ldots, s_{2r}, s_{2r+1}}(\nn, T) 
&\equiv \exists T_{0}, \ldots, T_{2r}\  (T=T_{0}, \ldots, T_{2r})  \wedge\\
& \displaystyle \bigwedge_{0\leq i\leq r} \text{Right}_{S, s_{2i}, s_{2i+1}}(\nn, T_{2i})
\wedge \displaystyle \bigwedge_{0\leq i< r} \text{Left}_{S, s_{2i+1}, s_{2i+2}}(\nn, T_{2i+1})
\end{array}
$$
with the predicate 
\begin{equation}
\label{eq:RR} 
\gamma_{RR}(s_{0}, s_{1},  \ldots, s_{2r}, s_{2r}, s_{2r+1}, r): 2r\leq K
\end{equation}
\begin{center} \fbox{$2r+1$ bounces} \end{center} 
$$
\begin{array}{ll}
\text{RL}_{S, s_{0}, s_{1}, \ldots, s_{2r+1}, s_{2r+2}}(\nn, T) 
& \equiv \exists T_{0}, \ldots, T_{2r+1} \ (T=T_{0}, \ldots, T_{2r+1}) \wedge  \\
&  \displaystyle \bigwedge_{0\leq i\leq   r} \text{Right}_{S, s_{2i}, s_{2i+1}}(\nn, T_{2i})
\wedge \displaystyle \bigwedge_{0\leq i\leq  r} \text{Left}_{S, s_{2i+1}, s_{2i+2}}(\nn, T_{2i+1})\\
\end{array}
$$
where $s_{0},  s_{1}, \ldots, s_{2r+1}, s_{2r+2}$ with the predicate 
\begin{equation}
\label{eq:RL} 
\gamma_{RL}(s_{0}, s_{1}, \ldots,  s_{2r+1}, s_{2r+2},r): 2r +1\leq K 
\end{equation}
\begin{center} \fbox{$2r+1$ bounces }\end{center} 
$$
\begin{array}{ll}
\text{LR}_{S, s_{0}, s_{1}, \ldots,  s_{2r+1}, s_{2r+2}}(\nn, T) 
& \equiv \exists T_{0}, \ldots, T_{2r+1} \ (T=T_{0}, \ldots, T_{2r+1})  \wedge\\
&\displaystyle \bigwedge_{0\leq i\leq  r} \text{Left}_{S, s_{2i}, s_{2i+1}}(\nn, T_{2i}). \wedge
 \displaystyle \bigwedge_{0\leq i\leq r} \text{Right}_{S, s_{2i+1}, s_{2i +2}}(\nn, T_{2i+1})
\end{array}
$$
where $s_{0},  s_{1}, \ldots, s_{2r+1}, s_{2r+2}$ satisfy the predicate 
\begin{equation}
\label{eq:LR} 
\gamma_{LR}(s_{0}, s_{1}, \ldots, , s_{2r+1}, s_{2r+2},r): 2r\leq K
\end{equation}
\begin{center} \fbox{$2r$ bounces} \end{center} 
$$
\begin{array}{ll}
\text{LL}_{S, s_{0}, s_{1}, \ldots, , s_{2r}, s_{2r+1}}(\nn, T) 
&\equiv \exists T_{0}, \ldots, T_{2r} \ (T=T_{0}, \ldots, T_{2r})  \wedge\\
&\displaystyle \bigwedge_{0\leq i\leq r} \text{Left}_{S, s_{2i}, s_{2i+1}}(\nn, T_{2i})
\wedge \displaystyle \bigwedge_{0\leq i< r} \text{Right}_{S, s_{2i+1}, s_{2i+2}}(\nn, T_{2i+1})
\end{array} $$
with  the predicate 
\begin{equation}
\label{eq:LL} 
\gamma_{LL}(s_{0}, s_{1},  \ldots, , s_{2r}, s_{2r+1},r): 2r\leq K \end{equation}
We let $\Gamma_{RR, r}, \Gamma_{RL,r}, \Gamma_{LR,r}, \Gamma_{LL,r}$ denote the finite set of predicates of the form
\ref{eq:RR}, \ref{eq:RL}, \ref{eq:LR}, \ref{eq:LL} respectively.
\begin{lemma}
 \label{le:runs} Consider a subset $S\subseteq Q$  and an integer $\kk>1$. 
 There exists a first-order formula $\text{\em Run}_{S}(N,s,s',p,p',T, \kk)$ that expresses the fact that 
  configuration $(s',p')$ is reachable, if ever, from configuration  $(s,p)$ at time $T$  by making at most $\kk$ traversals
 and without ever visiting a state in $S$
	except possibly for the first and last states of the run. 
	
	For  $K$  at least twice the number of states of the automaton, 
there  exists  a first-order formula $\text{\em Race}(N,s, p,T, K)$ 
that expresses the fact that $T$ is the minimum  time  for 
reaching from configuration $(s,p)$, if ever,  a  configuration in some  broadcasting state.

For  $K$  at least twice the number of states of the automaton, 
there  exists  a first-order formula \text{\em Mute}(N,s,p,K) that expresses the fact that 
no broadcasting configuration is ever reachable from configuration $(s,p)$.

\end{lemma}
\begin{proof}
The predicate $\text{Run}_{S}(N,s,s',p,p',T, \kk)$ is the disjunction 
of  the predicates  in the three cases below.
 Furthermore, for a fixed value of $T$ there is at most one pair
$(s',p')$ and at most one  predicate among the individual predicates in the expression of 
$\text{ Run}_{S}(N,s,s',$ $p,p',T, \kk)$ which is satisfied by the quintuple $N,s,s',p,p',T$

\fbox{Case 1: no rebound}
$$
\text{Run0}_{S}(N,s,s',p,p',T, \kk)\equiv \text{ Reach}_{S}(N,s,s',p,p',T) 
$$

\fbox{Case 2: one rebound}
$$
\begin{array}{l}
\text{Run1}_{S}(N,s,s',p,p',T, \kk)\equiv\\\
\displaystyle \bigvee_{t\in Q} \exists T_{1}, T_{2}\ (T = T_{1}+ T_{2} ) \wedge \text{ Reach}_{S}(N,s,t,p,\nn+1,T_{1})  \wedge \text{ Reach}_{S}(N,t,s',\nn+1,p',T_{2})) \vee\\
 \displaystyle \bigvee_{t\in Q}  \exists T_{1}, T_{2}\ (T = T_{1}+ T_{2}) \wedge \text{ Reach}_{S}(N,s,t,p,0,T_{1})  \wedge \text{ Reach}_{S}(N,t, s',0,p', T_{2}))\\
\end{array}
$$
\fbox{Case 3: more than one rebound}

\begin{figure}[H]
\begin{center}
\begin{tikzpicture}[scale=0.75]
\draw[dashed] (4,5) -- (0,4);
\draw (0,4) -- (5,3)-- (0,2)-- (5,1);
\draw (0,4) -- (5,3)-- (0,2)-- (5,1);
\draw[dashed] (5,1) -- (3,0);
\draw (0,5) -- (0,0);
\draw (5,5) -- (5,0);
\node at (0, 5.6) {$0$};
\node at (5, 5.6) {$\nn+1$};
\node at (4,5.2) {$s$};
\node at (-0.5, 4) {$s_{0}$};
%
\node at (5.5, 3) {$s_{1}$};
%
\node at (-0.5, 2) {$s_{2}$};
%
\node at (5.5, 1) {$s_{3}$};
\node at (3,-0.2) {$s'$};
 \end{tikzpicture}
\end{center}
\caption{Illustration of RunRR}
\label{fig:RR}
\end{figure}
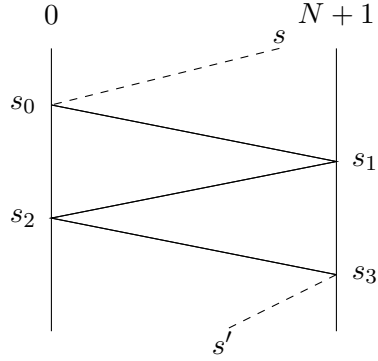
$$
\begin{array}{l}
\text{RunRR}_{S}(N,s,s',p,p',T, \kk)\equiv \displaystyle 
\bigvee_{1\leq 2r\leq K} \bigvee_{
\gamma\in \Gamma_{RR,r}} \\
 \text{ Reach}_{S}(N,s,s_{0},p,0,T_{1})   \wedge \text{ RR}_{S, s_{0}, s_{1}, \ldots,  s_{2r}, s_{2r+1}}(\nn, T_{2}) 
\wedge \text{ Reach}_{S}(N,s_{2r+1},s',\nn+1,p',T_{3}))\\
\end{array}
$$


%
\begin{figure}[H]
\begin{center}
\begin{tikzpicture}[scale=0.75]
\draw[dashed] (4,5) -- (0,4);
\draw (0,4) -- (5,3)-- (0,2)-- (5,1)--(0,0) ;
\draw[dashed] (0,0) -- (3,-1);
\draw (0,5) -- (0,-1);
\draw (5,5) -- (5,-1);
\node at (0, 5.6) {$0$};
\node at (5, 5.6) {$\nn+1$};
\node at (4,5.2) {$s$};
\node at (-0.5, 4) {$s_{0}$};
%
\node at (5.5, 3) {$s_{1}$};
%
\node at (-0.5, 2) {$s_{2}$};
%
\node at (5.5, 1) {$s_{3}$};
%
\node at (-0.5, 0) {$s_{4}$};
\node at (3,-1.2) {$s'$};
 \end{tikzpicture}
\end{center}
\caption{Illustration of RunRL}
\label{fig:RL}
\end{figure}
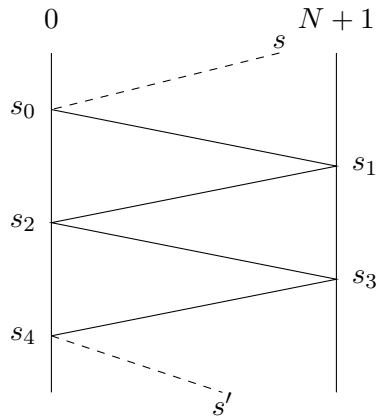
$$
\begin{array}{l}
\text{RunRL}_{S}(N,s,s',p,p',T, \kk)\equiv \displaystyle 
\bigvee_{1\leq 2r+1\leq K} \bigvee_{
\gamma\in \Gamma_{RL,r} } \\
\text{ Reach}_{S}(N,s,s_{0},p,0,T_{1})   \wedge \text{ RL}_{S, s_{0},  s_{1}, \ldots,  s_{2r}, s_{2r+2}}(\nn, T_{2}) 
\wedge \text{ Reach}_{S}(N,s_{2r+2},s',0,p',T_{3}))\\
\end{array}
$$
%
%


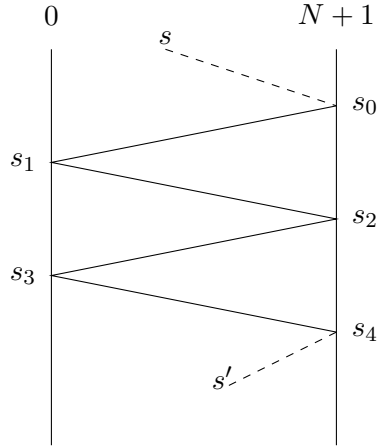
\begin{figure}[H]
\begin{center}
\begin{tikzpicture}[scale=0.75]
\draw[dashed] (2,6) -- (5,5);
\draw (5,5) -- (0,4)-- (5,3)-- (0,2)--(5,1) ;
\draw[dashed] (5,1) -- (3,0);
\draw (0,6) -- (0,-1);
\draw (5,6) -- (5,-1);
\node at (0, 6.6) {$0$};
\node at (5, 6.6) {$\nn+1$};
\node at (2, 6.2) {$s$};
\node at (-0.5, 4) {$s_{1}$};
%
\node at (5.5, 5) {$s_{0}$};
%
\node at (-0.5, 2) {$s_{3}$};
%
\node at (5.5, 3) {$s_{2}$};
%
%
\node at (5.5, 1) {$s_{4}$};
\node at (3,0.2) {$s'$};
 \end{tikzpicture}
\end{center}
\caption{Illustration of RunLR}
\label{fig:LR}
\end{figure}$$
\begin{array}{l}
\text{RunLR}_{S}(N,s,s',p,p',T, \kk)\equiv \displaystyle 
\bigvee_{1\leq 2r+1\leq K} \bigvee_{
\gamma\in \Gamma_{LR,r}} \\
\text{ Reach}_{S}(N,s,s_{0},p,\nn+1,T_{1})   \wedge \text{ LR}_{S, s_{0}, s_{1}, \ldots, s_{2r}, s_{2r+2}}(\nn, T_{2}) 
\wedge \text{ Reach}_{S}(N,s_{2r+2},s',\nn+1,p',T_{3}))\\
\end{array}
$$
%

\begin{figure}[H]
\begin{center}
\begin{tikzpicture}[scale=0.75]
\draw[dashed] (2,6) -- (5,5);
\draw (5,5) -- (0,4)-- (5,3)-- (0,2);
\draw[dashed] (0,2) -- (3,1);
\draw (0,6) -- (0,0);
\draw (5,6) -- (5,0);
\node at (0, 6.6) {$0$};
\node at (5, 6.6) {$\nn+1$};
\node at (2,6.2) {$s$};
\node at (-0.5, 4) {$s_{1}$};
%
\node at (5.5, 5) {$s_{0}$};
%
\node at (5.5, 3) {$s_{2}$};
%
\node at (-0.5, 2) {$s_{3}$};
\node at (3,1) {$s'$};
 \end{tikzpicture}
\end{center}
\caption{Illustration of RunLL}
\label{fig:LL}
\end{figure}
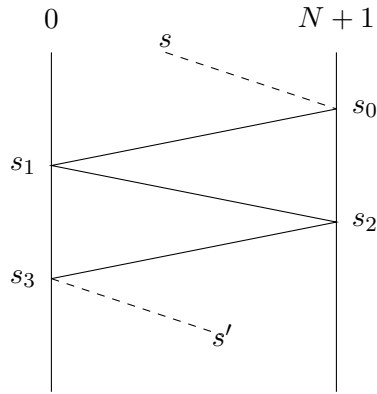
$$
\begin{array}{l}
\text{RunLL}_{S}(N,s,s',p,p',T, \kk)\equiv \displaystyle 
\bigvee_{1\leq 2r\leq K} \bigvee_{
\gamma\in \Gamma_{LL,r}} \\
 \text{ Reach}_{S}(N,s,s_{0},p,\nn+1,T_{1})   \wedge \text{ RR}_{S, s_{0}, s_{1}, \ldots,  s_{2r}, s_{2r+1}}(\nn, T_{2}) 
\wedge \text{ Reach}_{S}(N,s_{2r+1},s',0,p',T_{3}))
\end{array}
$$
Now, we pass to the second claim. 
Let $B$ be the set of broadcasting states. we want to express the fact that starting from configuration 
$(s,p)$ the automaton reaches configuration $(s',p')$ where $s'$ is broadcasting and no other broadcasting 
configuration was visited in the mean time.
For any $\kk\geq  2 |Q|$,  the following says that there exists a broadcasting state which is  the earliest
reachable state  from  configuration $(s,p)$ and that  $T$ is the time it is reached.
\begin{equation}
\label{eq:broadcasting-bis}
\begin{array}{l}
\text{Race}(N,s,p,T,K)  \equiv \displaystyle \bigvee_{b\in B}\
\text{(there exists a broadcating state reached from) } (s,p)\\
\big( (\exists T_{b}, \exists p_{b}\  \text{ Run}_{ \{b\}}(N,s,b,  p,p_{b},T, \kk)) \\
\text{(which is more quickly reachable than all other reachable broadcasting state)}\\
\wedge \displaystyle \bigwedge_{c\in B} ((\exists T_{c}, \exists p_{c}\  \text{ Run}_{ \{c\}}(N,s,c, p, p_{c},T_{c}, \nn))
\rightarrow T\leq T_{c})\big)
\end{array}
\end{equation}
Finally, the following expresses the fact that in configuration $(s,p)$ the automaton cannot reach any of its broadcasting states.
\begin{equation}
\label{eq:broadcasting-bis}
\begin{array}{l}
\text{Mute}(N,s,p,K)  \equiv \displaystyle \bigwedge_{b\in B} 
\displaystyle   \neg \big( \exists T_{b} \exists p_{b} \displaystyle \text{ Run}_{ \{b\}}(N,s,b,  p,p_{b},T_{b}, \kk)
\end{array}
\end{equation}

\end{proof}
\section{Dynamic of the multiautomaton}
%

We extend the definition of recognition of an input. 
A \emph{position function} is a function which assigns  to every $\nn\in \N$  an $n$-tuple $(p_{1}(\nn), \ldots, p_{n}(\nn))\in \{0, \ldots, \nn+1\}^{n}$   whose graph $\{(\nn, (p_{1}(\nn), \ldots, p_{n}(\nn))\mid \nn\in \N \}$  is Presburger definable.
Now, we apply this definition to  multiautomata.
Given an $n$-tuple of states $\sigma \in Q_{1} \times \cdots \times Q_{n}$ and a 
position function $f$,
an input $\nn$ is $(\sigma, f)-$ recognized by $\+M$ if starting in configuration $(\sigma, f(\nn))$ 
it reaches a configuration where some automaton is in a broadcasting state.

\begin{lemma}
\label{le:new-recognition}
The set of inputs that are $(\sigma, f)-$ recognized  is regular. Furthermore, 
for  each $\tau \in Q_{1} \times \cdots \times Q_{n}$   there exists at most a 
position function $g$ 
such that each $(\sigma, f)-$ recognizable input $\nn$ taking $\sigma$ to $\tau$ 
ends in configuration  $(\tau, g(\nn))$. 
\end{lemma}

\begin{proof}

We superscribe all the predicates of the previous section, $\text{Reach}$, $\text{Trav}$, 
$\text{Run}$, etc by the index of the automaton it
refers to. Given a global configuration $(\sigma, \pi)$ the automata compete for the next broadcasting configuration.

We want to give an upper bound $H$ on the time between an arbitrary configuration and the next braodcasting configuration.
Because the multiautomaton is deterministic and is composed of finite automata, between  two such events, some automaton (actually any automaton which eventually enters some of its broadcasting states)  makes a number of 
traversals  less than twice the number of its states which is less than or equal to twice the
maximum   number of states of all $n$ automata. Thus for each automaton we apply the 
formulas Run, Race and Mute in lemma \ref{le:runs} with the integer $K$ set to this maximum.
Let  $G \nn$  for some
integer $G$ be a bound on the maximum 
time of a left or right traversal of all automata, see
expression \ref{eq:right-left}. Then  we can take $H = GK  \nn$. 
Consequently, for each automaton, the number of traversals  during 
between two consecutive broadcasting configurations is bounded by $GK$ because it requires at least $\nn$ units of time for a traversal.

The following formula expresses the fact that starting in configuration $(\sigma, \pi)$ the multiautomaton reaches the next broadcasting
configuration $(\sigma', \pi')$. It assumes $\emptyset\subset I\subseteq \{1, \ldots, n\}$
to be  the set of indices of the automata that can reach a broadcasting state. 
More precisely it says that all automata 
$\+A _{i}$, $i\not\in I$ can no longer reach a broadcasting configuration from configuration $(\sigma, \pi)$ 
and that the earliest time the other automata reach a broadcasting configuration is $T$. Additionally it defines the new global configuration
 after $T$ units of time.
\begin{equation}
\label{eq:heat}
\begin{array}{l}
\Phi(N,\sigma,  \sigma',\pi, \pi', I, G, K)\equiv \displaystyle \big( 
\bigwedge_{i\not\in I} \text{Mute}^{(i)}(N,\sigma_{i},\pi_{i}, K)\big) \wedge \exists T\
(\exists (T_{i})_{i\in I} \ (T= \min\{T_{i}\mid i\in I\})\\
\displaystyle \wedge \bigwedge_{i\in I}\text{Race}^{(i)}(N,\sigma_{i},\pi_{i},T_{i},K)) 
\wedge \big(\displaystyle \bigwedge_{1\leq i \leq n}  \text{Run}^{(i)}_{\emptyset}(N,\sigma_{i},  \sigma'_{i}, \pi_{i}, \pi'_{i},T,  G \kk\big)
\end{array}
\end{equation}

Let $\Theta$ be the finite set of predicates $\theta(I,\sigma')$ defined by
$$
\begin{array}{l}
\emptyset \subset I\subseteq \{1, \ldots, n\}, \quad 
\sigma'\in Q_{1} \times \cdots \times Q_{n}, \quad
\sigma'_{j}\in B_{j} \leftrightarrow j\in I
\end{array}
$$
Let $f$ be a first-order definable function which assigns $\pi\in \{0, \ldots, \nn+1\}^{n}$ to every 
$\nn\in \N$, then the language which is $(\sigma,f)$-recognized by the multiautomaton
 is the set of integers $\nn$ satisfying  the first-order predicate
$$
(\pi=f(\nn))\wedge  \displaystyle \bigvee_{\theta\in \Theta} \exists \pi'\ (\Phi(N,\sigma,  \sigma',\pi, \pi', I, G, K))
$$
Observe that since the multiautomaton is deterministic, for every fixed $\nn$ there exits a unique $\theta\in \Theta$ such that 
the previous formula holds for some unique $\pi'$. Then $\tau=\sigma'$ and 
the first-order function $g$ assigns $\pi'$ to $\nn$. 
\end{proof}

{\bf Proof of Theorem \ref{th:finite-number-of-messages} }

The computation goes through exactly $M$ communication steps. It starts in configuration $(\sigma^{(0)},f^{(0)}(\nn))$
where $\sigma^{(0)}$ is the vector of the $n$ initial states of the automata and $f^{(0)}(\nn)=\pi^{(0)}$ with all components 
equal to $0$. A computation on input $\nn$ until the $M$-th broadcast is of the form
$$
(\sigma^{(0)}, f^{(0)}(\nn)), (\sigma^{(1)}, f^{(1)}(\nn)), \ldots, (\sigma^{M}, f^{M}(\nn))
$$
where for $i=0, \ldots, M-1$ the pair $(\sigma^{(i)}, f^{(i)}), (\sigma^{(i+1)}, f^{(i+1)})$ satisfies the condition of 
lemma \ref{le:new-recognition}. Since there are finitely many different such sequences and since each 
$(\sigma^{(i)},f^{(i)})$-recognized subset is regular it suffices to observe that after the $M$-th dispatch,
the only active automaton 
is $\+A_{1}$

\hfill $\Box$

\end{document}